\documentclass{aa}
\usepackage{natbib}
\usepackage{graphicx}
\begin{document}

\title{Afterglow Light Curve Modulated by a Highly
Magnetized Millisecond Pulsar}

\author{Heon-Young Chang,
	Chang-Hwan Lee, \and Insu Yi
}

\offprints{Heon-Young Chang}

\institute{
Korea Institute for Advanced Study \\
207-43 Cheongryangri-dong Dongdaemun-gu, Seoul 130-012, Korea\\
\email{hyc@kias.re.kr, chlee@kias.re.kr, iyi@kias.re.kr}\\
}

\date{Received ; accepted }

\abstract{
We investigate consequences of a continuously 
energy-injecting central engine of  gamma-ray burst (GRB) afterglow 
emission, assuming that a highly magnetized pulsar 
is left beaming in the core of a GRB progenitor.
Beaming and continuous energy-injection are natural consequences of 
the pulsar origin of GRB afterglows. 
Whereas previous studies have considered continuous 
energy-injection from a new-born pulsar to interpret the deviation of 
afterglow light curves of  GRBs from those with the
simple power law behavior, a beaming effect,
which is one of the most important aspects of pulsar emissions, 
is ignored in earlier investigations. 
We explicitly include the beaming effect and consider 
a change of the beaming with time due to a dynamical evolution of 
a new-born pulsar.
We show that the magnitude of the  afterglow from this 
fireball indeed first decreases with time, subsequently rises, 
and declines again. One of the most peculiar optical afterglows 
light curve of GRB 970508 can be accounted for by  continuous
energy injection with beaming due to a highly magnetized new-born
pulsar. 
We discuss implications on such observational evidence for a pulsar.
}

\authorrunning{Chang, Lee, \& Yi}
\titlerunning{Afterglow Light Curve Modulated by
a Highly Magnetized Millisecond Pulsar}
\maketitle

\section{Introduction}

Gamma-ray bursts (GRBs) are widely accepted to be produced when fast-moving, 
relativistic shells ejected from a central source in a relatively 
short period collide with slowly moving, yet relativistic
shells that were ejected at an earlier time 
(Rees \& ${\rm M\acute{e}sz\acute{a}ros }$ 1994; 
Paczy${\rm \acute{n}}$ski \& Xu 1994; Kobayashi et al. 1997;
Daigne \& Mochkovitch 1998, 2000).
In connection with the so-called internal shock model, the external shock
model also prevails as a possible origin of the GRB afterglows.
In the external shock model the relativistic material is 
assumed to be decelerated via interactions with the surrounding medium.
A shock wave results in heating the ambient matter to relativistic 
temperatures, and emitting photons in longer wave 
lengths (Rees \& ${\rm M\acute{e}sz\acute{a}ros }$ 1992; 
${\rm M\acute{e}sz\acute{a}ros }$ \& Rees 1993; 
Paczy${\rm \acute{n}}$ski \& Rhoads 1993; Sari et al. 1996; 
Vietri 1997).
GRBs and their afterglows seem to result from 
the dissipation of bulk energy in the relativistic outflows, which
are in the form of a narrow beam rather than a spherical shell.
 
Even though the origin of the observed GRBs are still unknown, from 
the observations of several GRB afterglows the evidence of beamed GRBs has 
been accumulated 
(Sari et al. 1999; Halpern et al. 1999; Rhoads 1999).
There are several works on models for 
the geometry of GRBs (e.g., Chang \& Yi 2001 and references therein) 
and their environments (e.g., Scalo \& Wheeler 2001 and references therein). 
Much of the current research on GRBs 
is aimed at determining the nature and the origin of the central engine 
(Duncan \& Thompson 1992; Narayan et al. 1992; Woosley 1993;
Katz 1994; Usov 1992, 1994a; Shaviv \& Dar 1995;
${\rm M\acute{e}sz\acute{a}ros }$ \& Rees 1997b; Yi \& Blackman 1997;
Blackman \& Yi 1998; Paczy${\rm \acute{n}}$ski 1998;
MacFadyen \& Woosley 1999; Portegies Zwart et al. 1999; Li 2000;
Wheeler et al. 2000; Zhang \& Fryer 2001).
Although the simple cosmological fireball afterglow model is in 
a good agreement with the observed light curves of a
power law decay (e.g., Wijers et al. 1997; Waxman 1997a,b), 
the afterglow of GRB970228 observed by HST, for instance, shows a deviation
from the simple power law behavior (Fruchter et al. 1997).
Several works have been done to further investigate 
more subtle effects that can change the afterglow 
characteristics (Katz \& Piran 1997; 
${\rm M\acute{e}sz\acute{a}ros }$ \& Rees 1997a, 1999; Rhoads 1999;
Berger et al. 2000; Dai \& Lu 2000; Kumar \& Panaitescu 2000a, b;
Panaitescu \& Kumar 2000; Dai \& Lu 2001).  

Most  fireball models assume that the energy 
injection into the fireball occurs in a short period of time
compared with the lifetime of the afterglows  
(Rees \& ${\rm M\acute{e}sz\acute{a}ros }$ 1998;
Kumar \& Piran 2000; Sari \& ${\rm M\acute{e}sz\acute{a}ros }$ 2000).
However, in some types of central engines, 
such as a fast rotating new-born pulsar with the strong magnetic 
field (magnetar), a significant energy 
input into the fireball may in principle continue for  
a significantly longer timescale, and
accordingly the temporal decay of the afterglow will be slower. 
Hence, it is worthwhile to investigate a continuously fed fireball in 
more details as a probe of the central engine of GRBs. 
We here consider the  central engine that emits both 
an initial impulsive energy input $E_{\rm imp}$ and
a continuous power. In fact, recently there were such attempts to
provide an explanation for the deviation of the afterglow light curve
from the simple power law (Dai \& Lu 1998a, b; 
Zhang \& ${\rm M\acute{e}sz\acute{a}ros }$ 2001a, b).
Even though previous studies  have considered continuous 
injection from a highly magnetized millisecond pulsar 
to interpret the deviation of 
afterglow light curves of some GRBs from the power law behavior,
one of the most important aspects of pulsar emissions is ignored, that is,
beaming. Beaming and continuous powering are the clear consequences of the 
pulsar origin of the GRB afterglows.  
In this Letter, this feature is included and a change of 
the beaming due to a dynamical evolution of the new born pulsar 
is explicitly taken into account.
We find that the deviations 
suggested by previous studies indeed occur, but that the shape of
the light curve is significantly modified due to a beaming effect.
If such observational evidence for beaming is found, 
the corresponding pulsar origin for
GRBs would imply that activities resulting in too massive objects 
to be a neutron star might be ruled out as a central engine of GRBs. 
Therefore, the question of whether or not
a signature of beaming can be found in the afterglows 
thus extremely important.

\section{Afterglows Continuously Powered by Pulsar}

The total luminosity emitted from a  young millisecond pulsar (MSP) 
has two important terms: an electromagnetic (EM) radiation term 
and a gravitational wave (GW) radiation term. 
Given that the spin-down is mainly due to electromagnetic dipolar 
radiation and to gravitational wave radiation, the spin-down law is
given by
\begin{eqnarray}
-I \Omega \dot{\Omega}=\frac{B_p^2 R^6 \Omega^4\sin^2\alpha}{6 c^3}
+\frac{32 G \epsilon^2 I^2 \Omega^6}{5 c^5}, \label{eq:zero}
\end{eqnarray}
where $\Omega$ and $\dot{\Omega}$ are the angular frequency and 
its time derivative, $B_p$ is the dipolar field strength at the 
magnetic poles, $\alpha$ is the angle between spin axis and the
magnetic field axis, $I$ is the moment of inertia, $R$ is the pulsar radius, 
and $\epsilon$ is the ellipticity of the compact star 
(Shapiro \& Teukolsky 1983).
When an MSP rotates with spin frequency larger than a certain critical 
frequency, $\Omega_{\rm crit}$,
nonaxisymmetric secular instability drives the pulsar into 
nonaxisymmetric configuration with nonzero quadrupole moment. 
This is the case where
the GW luminosity dominates. In this 
case, the spin-down of the MSP occurs on a 
timescale $\tau_{\rm gw}=I_* \Omega^2_*/2 L_{\rm gw} \sim 3 \times 10^{-3} 
\epsilon^{-2} I_{45} \Omega^{-4}_{4}$~s, where $I_*=I_{45} 
10^{45}~ {\rm g~cm^2}$ is the MSP moment of inertia, 
$\Omega_*=\Omega_{4} 10^4~{\rm s^{-1}}$ is the MSP spin frequency, 
and we have made use of the gravitational wave energy-loss rate, 
$L_{\rm gw}=32 G \epsilon^2 I_*^2 \Omega_*^6/5 c^5$. 
When an MSP rotates with spin frequency smaller than the certain critical 
frequency, the spin-down due 
to the electromagnetic dipole radiation gives a 
timescale $\tau_{\rm em} \sim I_* \Omega_*^2/2 L_{\rm em} \sim 3 
\times 10^2 I_{45} B_{15}^{-2} R^{-6}_{6} \Omega^{-2}_{4}~{\rm s}$, 
where $R_*=R_{6}10^6$ cm is the MSP radius, $B_*=B_{15} 10^{15}$ 
G is the MSP dipole magnetic field, and we have assumed the 
electromagnetic dipole energy-loss rate, 
$L_{\rm em}=2 \mu^2_* \Omega^4_*/3 c^3$, where $\mu_*=B_* R_*^3$ is 
the electromagnetic dipole moment.  
The decay solution $\Omega (T)$ as a function of time $T$
includes both EM and GW emissions,
but the energy input into the fireball is due to the EM dipolar 
emission $L_{\rm em}$ alone. 
Provided that the spin-down at a time is dominated by
one or the other loss term, one can obtain approximate solutions.
When GW radiation losses dominate the spin-down,
we have  $\Omega=\Omega_{0,{\rm gw}}(1+T/\tau_{\rm gw})^{-1/4}$, 
or approximately $\Omega=\Omega_{0,{\rm gw}}$
for $T \ll \tau_{\rm gw}$, 
and $\Omega=\Omega_{0,{\rm gw}}(T/\tau_{\rm gw})^{-1/4}$
for $T \gg \tau_{\rm gw}$, 
where $\Omega_{0,{\rm gw}}$ being the initial spin frequency. 
When EM dipolar radiation losses dominate the spin-down, we have 
$\Omega=\Omega_{0,{\rm em}}(1+T/\tau_{\rm em})^{-1/2}$, or 
approximately  $\Omega=\Omega_{0,{\rm em}}$
for $T \ll \tau_{\rm em}$, 
and $\Omega=\Omega_{0,{\rm em}}(T/\tau_{\rm em})^{-1/2}$
for $T \gg \tau_{\rm em}$,  
where $\Omega_{0,{\rm em}}$ being the initial spin frequency. 
Assuming $\Omega_{\rm crit} > \Omega_{\rm dynamo}$, 
where $ \Omega_{\rm dynamo}$ is the spin required for dynamo action, 
there are two types of MSPs: a supercritical strong field rotator (SPS) 
with $\Omega > \Omega_{\rm crit} > \Omega_{\rm dynamo}$, and 
a subcritical strong field rotator (SBS) 
with $\Omega_{\rm crit} > \Omega > \Omega_{\rm dynamo}$. 
These two classes may give the bimodal distribution of short and long 
pulsar-induced GRBs (Yi \& Blackman 1998).
For an SPS pulsar, GW spin-down is important in initial
period and promptly below $\Omega_{\rm crit}$.
As soon as  $\Omega < \Omega_{\rm crit}$
the spin-down becomes dominated by the EM losses. 
On the other hand, for an SBS pulsar, the spin-down is always
dominated by EM regime. 

The differential energy conservation relation 
for the self-similar blast wave can be written 
as $dE/dt = {\cal L}_0(t/t_0)^{q'} - \kappa' (E/t)$, where $E$ 
and $t$ are the energy and time measured in the fixed frame  
and $q'$ and $\kappa'$ are constants (Cohen \& Piran 1999). 
The first term  denotes the continuous luminosity injection, 
and the second term takes into account radiative energy losses 
in the blast wave. For $t>t_0$, the bulk Lorentz factor of 
the fireball scales with time as $\Gamma^2 \propto t^{-m}$, 
with $m$ and $\kappa'$ related by $\kappa'=m-3$ (Cohen et al. 1998).
If $m=3$, it corresponds to the adiabatic case (Blandford \& McKee 1976).
In the observer frame, the time $T$ is related to the fixed 
frame $t$  by $dT=(1-\beta)dt \simeq dt/ 2\Gamma^2$, 
and $T=\int^t_0 (2 \Gamma^2)^{-1} dt \simeq t/[2(m+1)\Gamma^2]$ 
when $t \gg t_0$. The differential energy conservation relation 
in the observer frame is now given by
$dE/dT = L_0(T/T_0)^q - \kappa (E/T)$, and can be integrated as
\begin{eqnarray}
E=\frac{L_0}{\kappa + q +1}\biggl(\frac{T}{T_0}\biggr)^q T 
+ E_{\rm imp} \biggl(\frac{T}{T_0}\biggr)^{-\kappa}, T > T_0,\label{eq:one}
\end{eqnarray}
where 
$T_0=t_0/[2(m'+1)\Gamma^2]$, $m'$ being the self-similar index for $t<t_0$,
$L_0=2 \Gamma^2 {\cal L}_0$, $q=(q'-m)/(m+1)$, and $\kappa = \kappa'/(m+1)$.
Setting $T=T_0$, the total energy at the beginning of 
the self-similar expansion is the sum of two terms, 
$E_0= L_0 T_0 /(\kappa + q+1)+E_{\rm imp}$. The first 
term is the accumulated energy from the continuous injection 
before the self-similar solution begins. The second term 
$E_{\rm imp}$ is the energy injected impulsively by the initial blast.

The total energy of the blast wave given by 
Eq.~\ref{eq:one} may be dominated either by the continuous injection 
term ($\propto T^{q+1}$) or by the initial impulsive term  
($\propto T^{- \kappa}$), subject both to the relative values of 
the two indices and to the values of $L_0$ and $E_{\rm imp}$ 
(Dai \& Lu 1998a, b; Zhang \& ${\rm M\acute{e}sz\acute{a}ros }$ 2001a).
One may classify three regimes according to
the relative values of the two indices as discussed by 
Zhang \& ${\rm M\acute{e}sz\acute{a}ros }$ (2001a).
We are interested in the case where  $q> -1-\kappa$, since otherwise 
a pulsar signature is no longer observable even if there is 
a pulsar in the central engine. 
If  $q> -1-\kappa$, the first term in Eq.~\ref{eq:one} 
will eventually dominate over the second term after a critical $T_c$. 
The injection-dominated regime begins at a critical time $T_c$ defined
by equating the injection and energy-loss terms  in Eq.~\ref{eq:one},
\begin{eqnarray}
T_c={\rm Max}\biggl\{ 1,\biggl[(\kappa+q+1)\frac{E_{\rm imp}}{L_0 T_0} 
\biggr]^{1/(\kappa+q+1)} \biggr\}. \label{eq:two}
\end{eqnarray}
If initially the impulsive term dominates, 
the critical time $T_c$  could be much longer 
than $T_0$, depending on the ratio of $E_{\rm imp}$ and $L_0 T_0$.
The case where $T_c \geq T_0$ ensures that a self-similar 
solution has already formed when the continuous injection term dominates. 
Besides the condition of $q> -1-\kappa$, for detecting the signature of 
a pulsar the critical time $T_c$ should be greater than  
$\tau_{\rm em}$ or $\tau_{\rm gw}$ as discussed below. 

To obtain the temporal decay index of the afterglow light curve
for which the MSP is responsible
we adopt the cylindrical geometry instead of the spherical geometry,
which may accommodate elongated beaming configurations. 
The rotational axis of the MSP coincides with the $z$-axis of the geometry.
For a fireball blastwave decelerated by a homogeneous external medium with 
particle number density $n$, the energy conservation equation 
at time $t=r/c$ is given by
\begin{eqnarray}
E=2 \pi R^2 r n m_p c^2 \Gamma^2= 2 \pi R^2 c t n m_p c^2 
\Gamma^2,&t>t_0, \label{eq:three}
\end{eqnarray}
where $r$ is the distance of the blastwave in $z$ direction,
$R$ is a radius of the beaming which can be characterized by that of 
the light cylinder, $c/ \Omega$, $\Omega$ being given by the MSP evolution,
and all other symbols have their usual meanings. 
One can work out the relationship between 
the temporal index $\alpha$ and the spectral index $\beta$, 
where $F_{\nu} \propto T^{\alpha} \nu^{\beta}$. 
To directly compare indices suggested by 
Zhang \& ${\rm M\acute{e}sz\acute{a}ros }$ (2001a),
we derived the scaling laws for the slow cooling regime
assuming that the reverse shock is mildly relativistic (Sari et al. 1998).
We derive the scaling law for $T\ll \tau_{\rm em}$ as an illustrating example.
The energy E  in Eq.~\ref{eq:one} should have the same time dependence 
as the first term on the right-hand side of Eq.~\ref{eq:zero}, 
giving $T^{q+1} \propto t \Gamma^2$, or $m=-q/(q+2)$. Since in general, 
$\Gamma \propto t^{-m/2} \propto r^{-m/2} \propto T^{-m/2(m+1)}$ and
$r \propto T^{1/(m+1)}$, the $\Gamma$ dependence on $T$ 
is given by  $\Gamma \propto T^{q/4}$. 
The relationship between 
the temporal index $\alpha$ and the spectral index $\beta$
can be obtained as below. 
For the forward shock, the synchrotron peak frequency 
$\nu^f_m \propto \Gamma^4 \propto T^{-2m/(m+1)} \propto T^{q}$,
the peak flux $F^f_{\nu_m} \propto T^3 \Gamma^8 \propto T^{(3-m)/(m+1)} 
\propto T^{3+2q}$. Thus, $\alpha^f=3+2q-q \beta=(2m \beta^f+3-m)/(1+m)$ 
since $T^{\alpha^f} \nu_m^{\beta^f} \propto T^{3+2q} \propto T^{(3-m)/(m+1)}$.
For the reverse shock, 
$\nu^r_m=\nu^f_m/\Gamma^2 \propto \Gamma^2 \propto T^{-m/(m+1)} \propto T^{q/2}$,
the peak flux $F^r_{\nu_m}=\Gamma F^f_{\nu_m} \propto T^3 \Gamma^9 
\propto T^{(6-3m)/2(m+1)} 
\propto T^{(12+9q)/4}$. 
Thus, $\alpha^r=(2 m \beta^r+6-3 m)/2(1+m)=(12+9q-2q\beta)/4$.
Similarly, one can derive the  temporal index for each regime. 
In Table 1 we have summarized temporal indices in various regimes for 
the forward shock and the reverse shock, respectively. 

If the continuous injection term becomes dominant over the impulsive 
term after $T_c$, the afterglow light curves rises after $T_c$ and steepen
again after some time, that is, about $\tau_{\rm em}$ or $\tau_{\rm gw}$. 
Therefore, there may be two types of afterglow patterns
for continuous injections according to a appropriate combination.
Zhang \& ${\rm M\acute{e}sz\acute{a}ros }$ (2001a) discussed
conditions which allow to detect a signature for a pulsar 
and concluded that physical parameters are consistent with those
of a magnetar in case of the afterglow features in case of GRB 000301c. 
In practice, however, $\tau_{\rm gw}$ is very short, can be even shorter 
than $T_c$ unless ambient matter density is very high, 
and is therefore unlikely to be observed.
At around $T_c$ and $\tau_{\rm em}$, 
$q$ changes -1 to 0 and 0 to -2, respectively. 
These scaling laws represent a change from the standard adiabatic 
case to an EM-loss dominated regime as shown above (see also
Zhang \& ${\rm M\acute{e}sz\acute{a}ros }$ 2001a). 
For the forward shock, the temporal decay index
changes around $T_c$ from  $3\beta^f/2$ to $3$ and returns to $3 \beta^f-3$ 
after $\tau_{\rm em}$.
For the reverse shock,  the temporal decay index
changes around $T_c$ from $3\beta^r/4-3/8$ to $3$ and 
returns to $3 \beta^r/2-15/4$ after $\tau_{\rm em}$.

\begin{table}
\begin{center}
\caption{Temporal index $\alpha$ for the
forward and the reverse shocks in various timescales. }
\vspace{2mm}
\begin{tabular}{cll}
\hline\hline
&$~~~~~~~~~~\alpha^f$&$~~~~~~~~~~\alpha^r$ \\ \hline \hline
$T \ll \tau_{\rm gw}$ & $ 3+2q-q\beta^f $     &$ (9q+12-2q \beta^r)/4  $\\
$T \gg \tau_{\rm gw}$ & $2+2q-\beta^f(q-1/2)$ &$ (18q+15-2(2q-1) \beta^r)/8$ \\ 
\hline
$T \ll \tau_{\rm em}$ & $3+2q-q\beta^f $      &$ (9q+12-2q \beta^r)/4 $\\
$T \gg \tau_{\rm em}$ & $1+2q-\beta^f(q-1)$   &$ (9q+3-2\beta^r(q-1))/4 $ \\
\hline \hline
\end{tabular}
\end{center}
\end{table}

\section{Discussions}

A continuous energy injection signature in the GRB afterglow light curve
may directly provide 
diagnostics about the nature of the injection as well as information 
on the GRB progenitor. 
Therefore, the question of whether or not
a  bump in the afterglow light curve is such an 
observational evidence for beaming from a pulsar
is thus extremely important. In this sense, the correct geometry
should be applied to the model calculation.
We have discussed the case where a pulsar is continuously injecting energy 
{\it cylindrically} rather than spherically. We suggest that a 
possible explanation for the deviation of the afterglow light curve
of GRB 970508 is due to a beaming from a central pulsar. The optical 
afterglow of GRB 970508 has been explained as evidence of gravitational 
lensing event (Loeb \& Perna 1998; Dado et al. 2001). 
So far the gravitational lensing
can only account for the fact that the GRB afterglow shows a rise
 and a fall.
We show in this Letter that with a correctly assumed geometry of 
energy injection the afterglow of GRB 970508 can be explained by 
a strongly magnetized fast-rotating pulsar with 
the continuous energy injection in a beam. 
In this calculation we ignore effects of the lateral expansion of
the jet which may occur in the later time of the jet evolution. 
The light curve should be modified and becomes that of the spherical
geometry case if the sideways expansion occurs in a timescale
comparable to those we discussed after a neutron star formed. 
For instance, if the jet is expanded much
faster at earlier stage than the light cone evolution, the cylindrical
geometry effect becomes less obvious. 

If the bump is caused by such a pulsar indeed, it puts constraints 
on the GRB progenitor models. That is, the corresponding pulsar origin for
GRBs would imply that activities resulting in too massive objects 
to be a neutron star, such as, neutron star mergers, black hole formation, 
should be ruled out as a central engine of GRBs.
During the birth of the neutron star, an initial fireball may occur through
electromagnetic processes (Usov 1992).
This has led to models in which GRBs are powered by rapidly spinning
compact objects with strong magnetic fields 
(Blackman \& Yi 1998; Blackman et al. 1996; 
Usov 1994b; Yi \& Blackman 1997, 1998).

\acknowledgements{
We are grateful to the referee, Robert Mochkovitch,  for useful 
comments and suggestions.
We are grateful to Ethan Vishniac for hospitality while
visiting Johns Hopkins University where this work began.}

\end{document}